# Psychological State Monitoring Methods Based on Multimodal Emotion Recognition

Ran Meng [a,1]
[a] *Jianghan University Mental Health Education Center, Wuhan, Hubei Province, 430056, China*

**Abstract.** The computer-Based Multimodal Emotion Recognition System needs to increase the accuracy and reliability of psychological state monitoring. A hierarchical framework was proposed in this paper to integrate facial images, speech signals and other physiological indicators of emotion in time and space for more complete analysis of human emotions. Synchronously sample data streams from webcams, microphones and wearable sensors, and then preprocess and extract features specific to each type of data. A high-end fusion architecture is adopted to combine early- and late-fusion methods and thus robustly integrate cross-modal information. Experimental validation in a real-world stress-response dataset showed good performance. The overall recognition accuracy and mean F1-score of the fused multimodal system are both about 91% and 0.88, respectively, which are superior to the best unimodal approach (face, 88% accuracy and F1-score 0.88). Temporal analysis shows that the groups for stress tracking are not uniformly adapted and resilient, and ablation experiments have determined which modules contribute significantly to the stability of the system. According to all kinds of quantitative evaluations of precision, recall, accuracy and F1 scores, the proposed method shows good balance and is practical and generally applicable. The research has provided an engineering-feasible, data-driven solution for precise emotion inference and offered methodological support for intelligent mental health assessment and human-centered system design.

**Keywords.** Speech Processing, Physiological Signal Analysis, Multimodal Fusion, Emotion Recognition, Psychological State Monitoring, Machine Learning, Real-Time Systems

## 1. Introduction

Monitor and evaluate psychological conditions in all parts of society for the purposes of mental health management, intelligent human-computer interaction, and safety systems. With the rapid development of affective computing, emotion recognition can now be used to assess the psychological well-being and mental health of people reliably in both clinical and non-clinical settings [1]. The current method of analysing emotions based on faces, speech and other changes in the body is quite accurate and is becoming more common [2]. However, given that human emotions are complex and multifaceted, expressed in various forms such as words and actions, most current emotion recognition models are struggling to handle all types and conditions of emotions in real life [3]. Such single-modality systems are prone to issues in practice, such as fluctuations in light, occlusion, noise or other behaviours by the subject, and consequently exhibit lower recognition accuracy and reduced stability [4].

[1] Corresponding Author: Ran Meng

Although single-modal systems are relatively simple to use and implement, they are inherently limited by problems such as signal degradation and contextual ambiguity; they also offer an incomplete expression of affective states [5]. For example, face-expression-based methods may have poor performance in low-light conditions or when faces are partially obstructed; speech-based emotion recognition is easily affected by noise in the environment or different accents; and although physiologically-based indicators are not as subjective, they also fail to reveal the whole picture of a person's emotions and moods [6]. Therefore, the latest studies have begun to focus on integrating multiple types of data, and it has been put forward that combining facial expressions, speech, etc., and other physiological signals can improve the comprehensiveness and reliability of emotion- and psychological-state detection [7].

Therefore, this paper puts forward a full-featured multimodal emotion recognition system that can accurately and reliably monitor people's emotional states. Combine face, speech and physiological characteristics in this way using deep-learning methods and data-fusion technology to achieve a high-accuracy, stable-temporal-series recognition system. The first is a design for a hierarchical multi-modal monitoring architecture; the second is a development of an end-to-end fusion model for emotion inference; and the third is systematic experimental verification using real-world data in a stress-response setting. The rest of this paper is organised as follows: Section 2 discusses relevant studies; Section 3 introduces the proposed system architecture; Section 4 describes the method for monitoring psychological states and conducts experiments; and finally, Section 5 provides a conclusion to this paper.

## 2. Related Work

### *2.1. Advances in Emotion Recognition and Multimodal Fusion*

Traditionally, the different forms of emotional expression were not systematically studied for this purpose, such as facial expressions, vocal cues and other physiological changes. Typically, the methods of facial analysis are based on geometric or appearance-based features for discrete emotion recognition, but they are often affected by changes in light and facial occlusion, as well as inter-individual differences [8]. Most speech-based emotion recognition methods obtain prosodic features of speech, such as pitch, speaking speed and other characteristics in the frequency spectrum, to identify emotion; however, factors such as environmental noise and variations among different speakers can reduce the accuracy of recognition [9]. Physiological indicators of different states of arousal and stress include fluctuations in heart rate variability and changes in electrodermal activity; however, the specific values obtained are also influenced by an individual's baseline level and their current environment.

Given the above constraints, recent progress in emotion recognition has focused on integrating multiple types of complementary data. At present, multimodal fusion technology is used to address the deficiencies of ambiguity and context dependency in single-modality methods. Early fusion joins features from various modalities at the input or in an intermediate stage, and late fusion merges the results of individual classifiers. High-end strategies based on deep learning can use adaptive or hierarchical structures to learn cross-modal relationships and modality-specific features at the same time. By integrating the three sources of information—images, sounds and bodies—the stability and generalisation capability of the system have been enhanced; thus, recognition accuracy has improved.

### *2.2. Methods for Psychological State Monitoring*

The scope of the field of psychology monitoring has expanded; previously, it was done through self-reporting and doctor-patient interviews, but now automated systems powered by machine learning and computation are being developed. The original evaluation method is suitable for some particular cases but is not feasible in all real-time situations. Recently, some systems have added wearable sensors, environmental microphones and computer vision to collect behavioural and physiological indicators in an unobtrusive manner for long-term tracking of changes in emotion and mental health.

However, the current solutions in most of the research have some deficiencies. Many systems are still restricted to a single type of data or employ a simple fusion method, and thus cannot handle the complexity of psychological phenomena or the dynamics of the natural environment. Static decision thresholds are less adaptable to the context and changes of life. Therefore, a comprehensive multimodal framework is urgently needed to accurately and consistently infer a person's mental state by integrating various sources of information and overcoming the shortcomings of any single signal.

## 3. Proposed Multimodal Emotion Recognition Framework

### *3.1. Data Acquisition and Preprocessing*

The proposed system continuously acquires multiple kinds of data, such as faces, speech and health. A high-definition camera is used to acquire facial information in a fixed Position for consistency. A high-quality microphone in the studio records the speech signal in real time, and a rich time-series data for paralinguistic feature extraction is obtained. Wearability sensors acquire physiological data, such as heart rate (HR) using photoplethysmography (PPG) and galvanic skin response (GSR) through electrodes, are non-invasive and suitable for prolonged observation.

All incoming signals are preprocessed to improve signal quality and reduce artifacts. Automatic face detection and landmark localisation are performed on the facial image, and then non-facial regions are suppressed. To handle changes in light and slight displacements of the body, histogram equalization and affine transformations are used for image normalization. A pipeline of denoising, silence removal and amplitude normalisation are used to process speech data and stabilise the audio intensity of all sessions. First of all, the band-pass filter is applied to reduce high-frequency noise and baseline drift in the physiological signals. Then, z-score normalisation is performed:

$$X_{\text{norm}} = \frac{X-\mu}{\sigma} \tag{1}$$

where $X$ is the raw signal, $\mu$ and $\sigma$ denote the sample mean and standard deviation, respectively. All modules in time are aligned through frame interpolation or window alignment to obtain a unified and temporally synchronous input for the following analysis.

### 3.2. Feature Extraction for Modalities

The features extracted by the different modes are not the same. Facial data have been represented by appearance-based descriptors and geometric configurations. Histogram of Oriented Gradients (HOG), Local Binary Patterns (LBP), and a set of 68-point facial landmarks are computed frame-by-frame. Finally, a current convolutional neural network maps these to expression action units (AUs) to capture fine-grained muscle movements.

Extract prosodic, spectral and voice-quality features from speech analysis. Melfrequency cepstral coefficients (MFCCs), pitch contours, jitter, shimmer and formant frequency trajectories are all sampled at 10 ms intervals to obtain micro-level emotional cues in vocal patterns. A Change in dynamic Pitch and energy are also associated with arousal.

Physiological signals have time-domain and frequency-domain features extracted from HR and GSR. Heart rate variability (HRV), root mean square of successive differences (RMSSD), and power spectral density components are all manifestations of the autonomic nervous system. GSR analysis finds the number and height of phasic peaks to determine sympathetic arousal. Standardise the feature-level representations of all modalities to unit variance and then concatenate them for joint modelling.

### 3.3. Fusion Strategy and Model Construction

The three proposed architectures are early fusion, late fusion, and an advanced hierarchical fusion. Early fusion is employed to combine normalised features from the face, speech and other body-related sources, then these features are fed into a deep feedforward network for end-to-end learning of cross-modal relations. This way uses the rich spatial-temporal features of multiple modalities and is suitable for end-to-end optimisation.

Late fusion can be used to combine independent modality-specific classifiers trained in different areas, and then a weighted-voting method can be employed to combine their results. Dynamically adjusting class probabilities in the system.

$$P_{\text{final}} = \sum_{m=1}^{M} w_m P_m \tag{2}$$

where $P_{\text{final}}$ is the decision vector, $M$ is the number of modalities, $P_m$ is the probability vector from the $m$-th modality, and $w_m$ are learned modality weights.

Our hierarchical fusion combines the two paradigms: first, early fusion is used for correlated features (e.g., audio-visual), and then aggregation with physiological-based predictions occurs at the decision level. The architecture of the above system is shown in Figure 1, and the entire process of multi-modal data acquisition and alignment, followed by high-level psychological state inference, is also illustrated here.

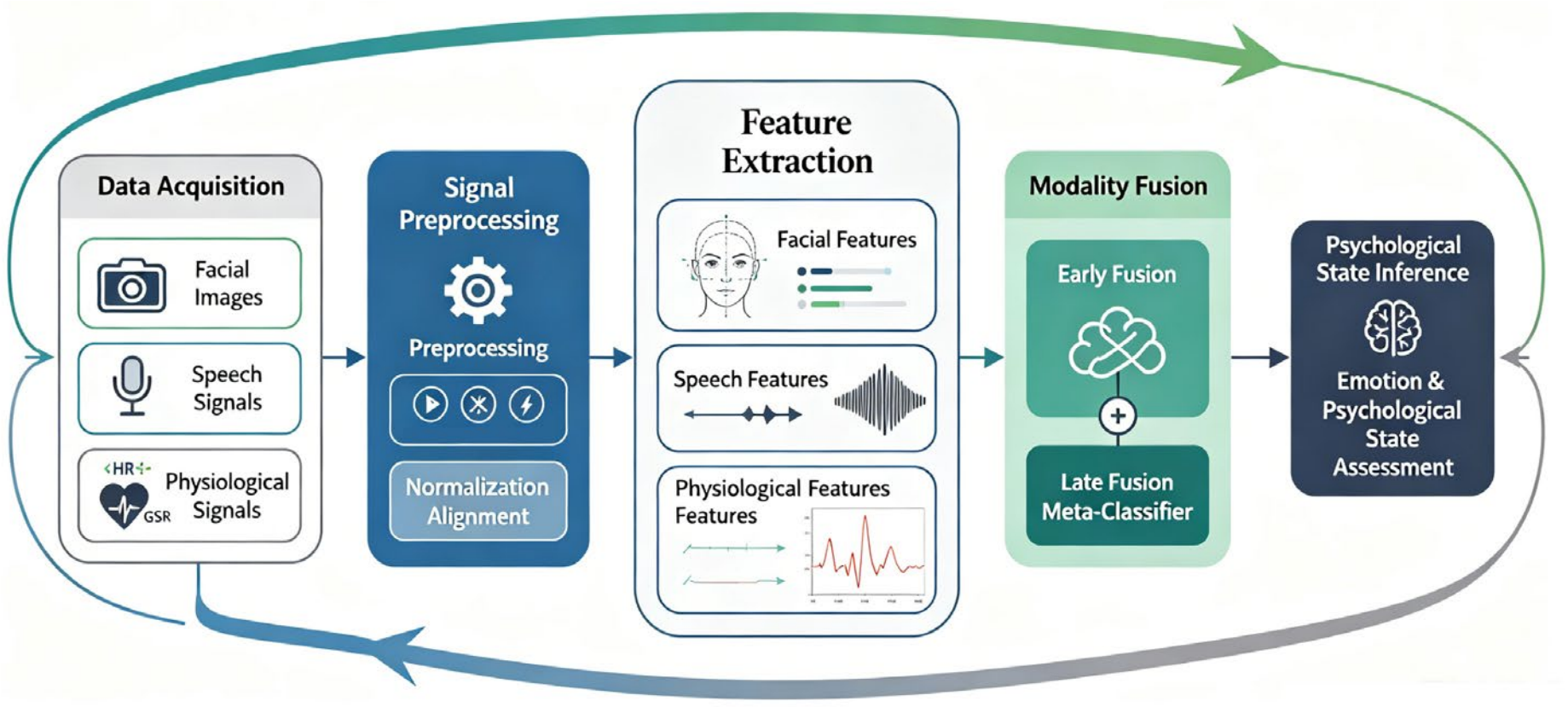

**Figure 1.** Multimodal Emotion and Psychological Monitoring: Workflow

## 4. Psychological State Monitoring and Experimental Evaluation

### *4.1. Hierarchical Monitoring Framework & Data Analysis*

A hierarchical framework has been introduced to handle the high-dimensional and multi-modal characteristics of emotion recognition in the psychological state monitoring system. The three parts of this system are: data, features and decisions; they work together to do their jobs. Collect the four channels of synchronized raw data from the data layer: facial images, speech signals, heart rate (HR), and galvanic skin response (GSR). Pre-processing and feature extraction routines are customized for all types of data at the feature level, including facial landmark localisation, MFCCs for speech, and HRV descriptors [10].

Fusion Strategies are applied in the decision layer to enhance both the robustness and accuracy. Early fusion merges low- and mid-level features from both modalities that are highly correlated in time, and a weighted late fusion approach is used to incorporate physiologically derived results into the decision-making process [11]. Several levels can be set to make use of the strengths of cross-modal synergy and increase robustness to partial data loss.

According to the above ablation experiments and importance analysis, the unimodal and multi-modal versions have varying levels of performance, as shown in Figure 2. The highest unimodal recognition accuracies were 88% for face, 81% for speech, 74% for heart rate and 66% for GSR. Facial data were by far the most influential single-modality contributors, and speech and physiological cues added specificity and discriminability to the fused model. Detection scores for the classes of positive, neutral and negative emotions in all three modalities were also shown to be relatively balanced, and differences in the contribution of the visual channel and other modalities were identified. The dumbbell plot is used to display the range and mean of each channel explicitly, and it can be seen that integrated multimodal fusion is necessary [12]. Therefore, the comprehensive mode of stable psychological monitoring is highly recommended by the above study [13].

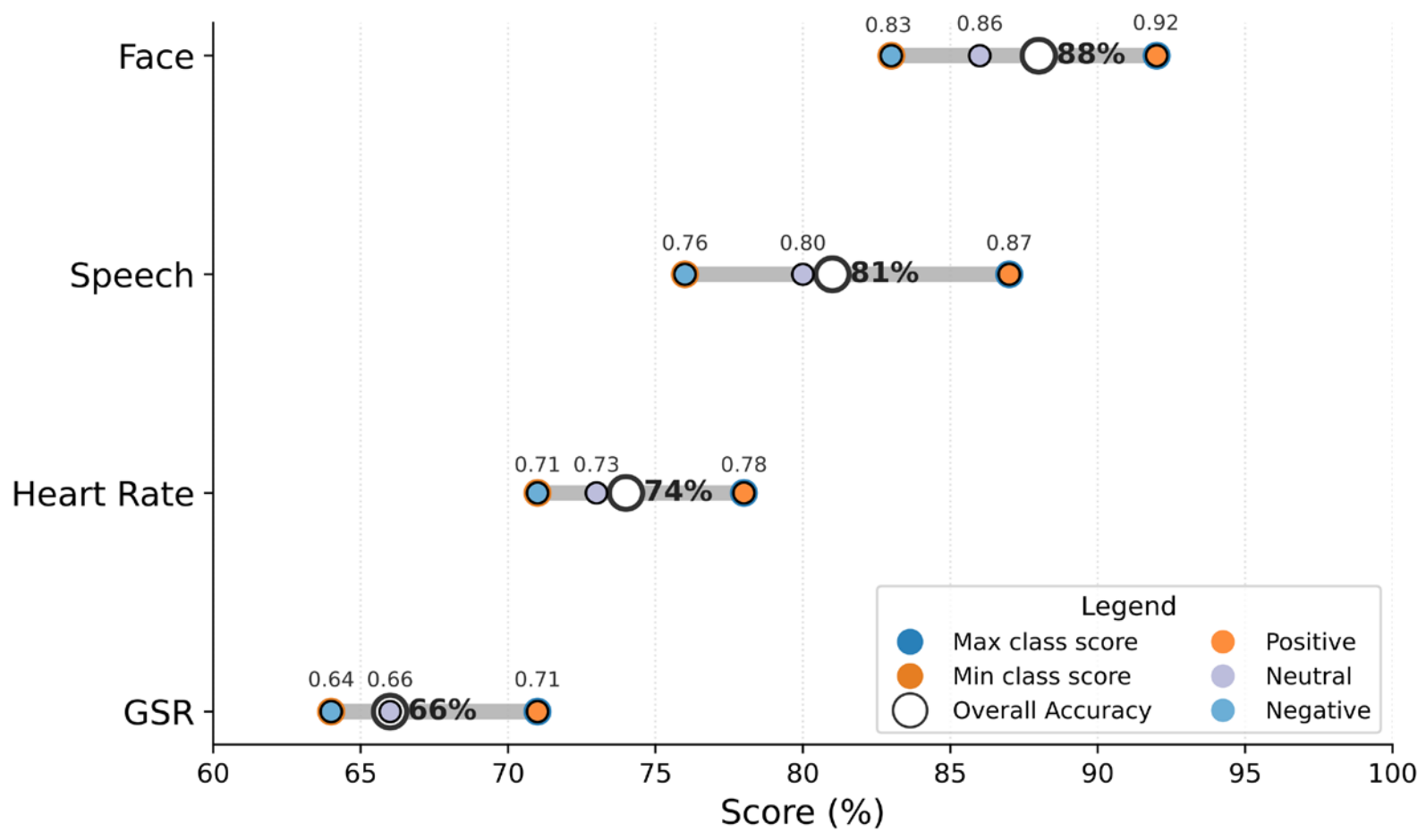

**Figure 2.** Contribution of Each Modality to Overall Emotion Recognition Accuracy.

### *4.2. Real-time Tracking & Group-level Results*

Multiple groups of participants were randomly divided into two categories: stress-challenge and relaxation, and then continuously recorded. At each sampling time, the system calculated and saved the emotional and stress indices for all subjects to obtain time-series data [14].

The group-level results shown in Figure 3 are clearly different in stress response and adaptation. Group A showed an early acute peak and then a gradual decline to 32 by the end of the window. Group C also recovered but had a slightly higher baseline. Groups B and D showed a relatively small range and a slower response speed; therefore, their curves were less steep. The various groups have different degrees of stress resistance and recovery abilities for individuals and groups. Continuous monitoring at a high temporal resolution can be used for both fine-grained within-subject analysis and stable between-group statistical tests [15]. Such precise, time-synchronised tracking can be used to study changes in psychological adjustment over time and is also suitable for research and applications in monitoring workplaces or clinics [16].

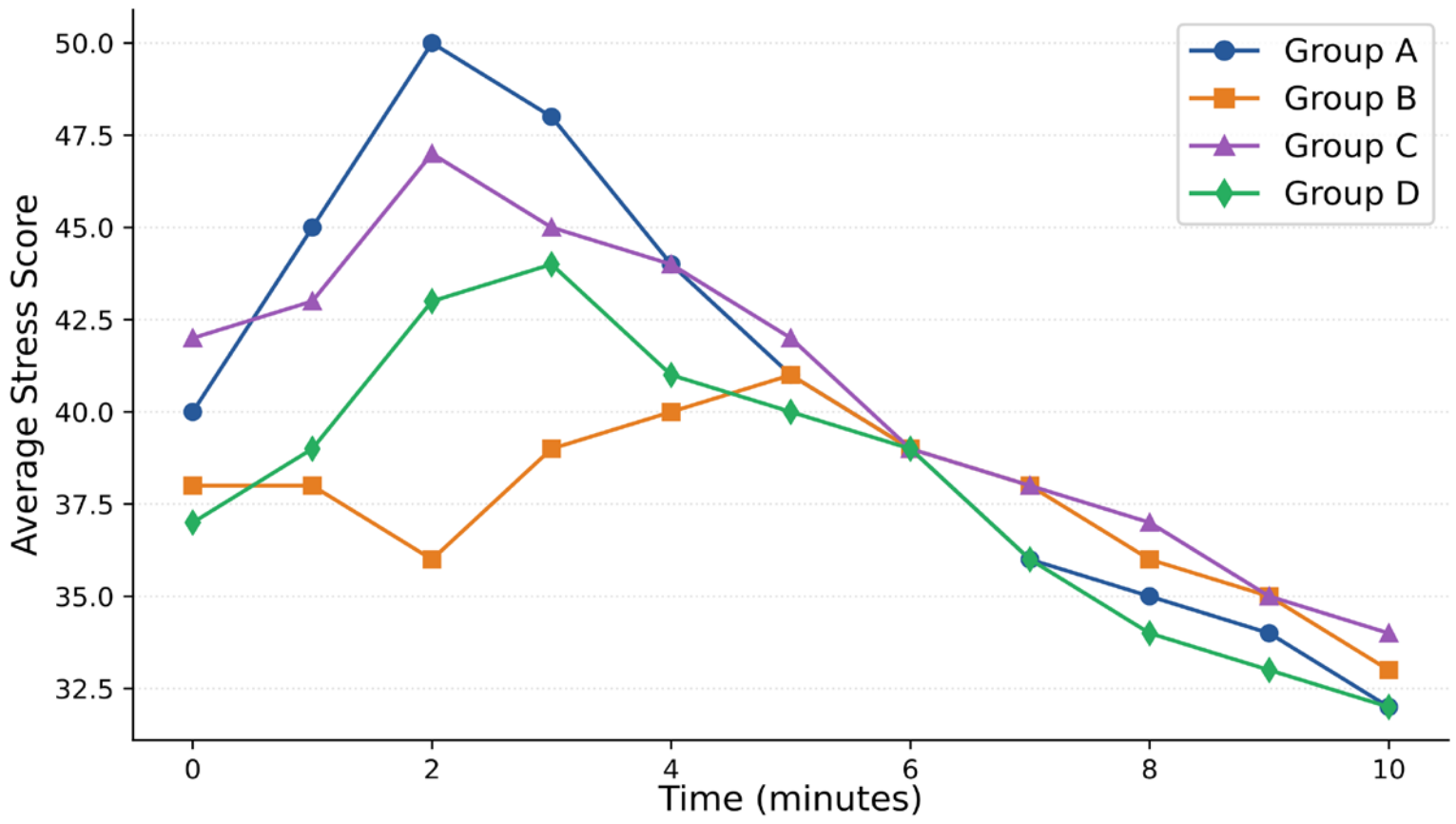

**Figure 3.** Temporal Changes of Average Psychological Stress Scores for four groups (A–D).

*4.3. System Implementation & Multi-metric Performance*

A modular, expandable framework was developed and tested for real-time streaming [17]. Precision, recall, accuracy, and F1-score were used to evaluate each modality and the fused system. As Figure 4 shows, the multimodal system achieves high, stable metrics: Face precision 0.89, recall 0.87, accuracy 0.88, F1-score 0.88; Speech metrics all exceed 0.80; Heart Rate and GSR yield average F1-scores of 0.74 and 0.66. Fusion elevates overall performance, with a mean F1-score of 0.77. Even after ablation of any modality, fusion performance remained above the best single modality, demonstrating robust multi-source integration [18]. The grouped bar plot and trend line illustrate the improved balance and robustness for emotion and psychological state recognition [19].

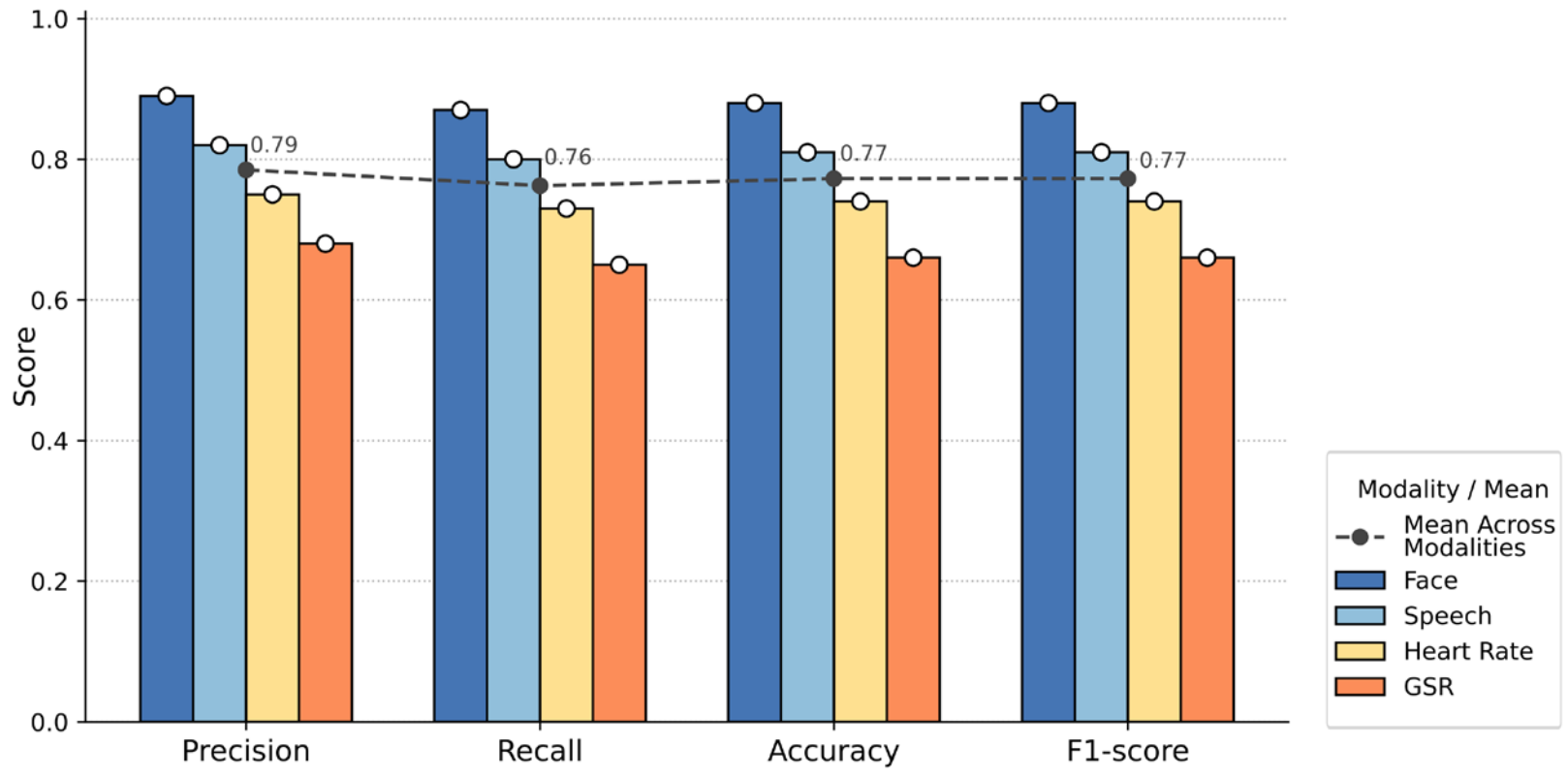

**Figure 4.** Multi-dimensional Emotion Recognition Performance for Face, Speech, Heart Rate, and GSR.

## 5. Conclusion

This paper proposes an all-encompassing, hierarchical system for multi-modal psychological state monitoring and emotion recognition based on face images, speech and physiological signals, etc., and applies special feature extraction and staged data fusion. According to the above experiments, the new system performs better than that of the unimodal baseline consistently in terms of recognition accuracy, and the combination of facial and speech features has brought significant improvements to this. The developed platform can monitor the psychological state of a group in real time and is therefore suitable for research and application; a combination of indicators is used to test how stable and broadly applicable the method is under changes and noise.

The above provisions still have some defects. The system is sensitive to hardware quality and sensor synchronisation, and its data collection may be confined to controlled environments; thus, it is not suitable for off-line use in the field. Additionally, differences in people's emotions and bodies cannot be fully accounted for by the current model's structure. Although the current ensemble strategies are good, they can be more adaptive or personalised in learning.

Future work will expand the platform to include more diverse groups of people in various circumstances and improve ecological validity. Add more sophisticated calibration and adaptive signal processing, and use a transparent deep learning model.

Finally, the combination of wearable and mobile sensing and continuous self-supervised learning will be used to build a large-scale, all-weather psychological monitoring system.